# Experimental Signatures of Topological Transport in Weyl semimetal candidate FeSi

R. Mantovan[1,a),b)], A. Bozhko[2], V. Zhurkin[2], A. Bogach[2], A. Khanas[3], S. Zarubin[3], A. Zenkevich[3], V. Glushkov[2,a),b)]

[1]*CNR-IMM Unit of Agrate Brianza, Via Olivetti 2, 20846 Agrate Brianza (MB), Italy*

[2]*Prokhorov General Physics Institute of Russian Academy of Sciences, 119991 Moscow, Russia*

[3]*Moscow Institute of Physics and Technology ((National research university)), 141701 Moscow Region, Russia*

a) *These authors contributed to this paper equally*

b) *Correspondence and requests should be addressed to V.Glushkov (glushkov@lt.gpi.ru) and R.Mantovan (roberto.mantovan@cnr.it)*

Iron monosilicide ε-FeSi has recently been predicted to host nontrivial topological states, yet experimental evidence remains scarce. Here we report transport signatures of Weyl semimetal behavior in polycrystalline ε-FeSi films (65 nm thick) grown by solid-state reaction of Fe on Si(100). Below 200 K, the anomalous Hall conductivity $\sigma_{xy}^{AHE}$ becomes temperature-independent ($\sigma_{xy}^{AHE} \approx 14$ μS/sq. ~ const ($\sigma_{xx}$)), unequivocally demonstrating an intrinsic anomalous Hall effect driven by Berry curvature. By comparing our data with the large set of data available in the literature for FeSi single crystals and polycrystalline/amorphous/crystalline FeSi thin films of thicknesses between 10 nm and 1 mm, we discover a universal scaling of $\rho_{xy}^{AHE} \sim \rho_{xx}^2$, which confirms the existence of surface-dominated electron transport. In this regime, the chiral anomaly manifests in both anisotropic longitudinal magnetoresistance and the planar Hall effect as well. The above observations establish ε-FeSi as a Weyl semimetal and allow to relate $\sigma_{xy}^{AHE}$ to a "quantized" Hall response estimating the effective Weyl-point separation as $(k_+^W - k_-^W)/(2\pi) \approx 0.36$. Our findings confirm the topological origin of electron transport in ε-FeSi and discover its potential as a new high temperature and noble metal-free Weyl semimetal.

**Introduction.** Recently, non-centrosymmetric *M*Si monosilicides (*M*=Fe, Co, Mn, Rh, Ru etc., space group 198, $P2_13$) have renewed keen interest among researchers after *ab initio* calculations predicted exotic massless fermionic excitations with nonzero Berry flux in CoSi [1] and unconventional chiral fermions and large topological Fermi arcs in RhSi [2]. Nontrivial topological entity of *M*Si was supported by novel gapless surface phonons at double-Weyl points reported for these crystalline materials (*M*=Fe, Co, Mn, Rh, Ru) [3] and proved experimentally for parity breaking FeSi single crystals with the help of inelastic x-ray scattering [4]. The above findings are well consistent with a general conclusion that all the space groups without inversion symmetry have a general trend for entering into topological phase of Weyl or nodal-line semimetal under gap collapsing [5].

While 3D topological insulating or Weyl semimetallic behavior in FeSi suggested in [6] was not directly proved by *ab initio* DFT calculations [7], inherent surface ferromagnetic-metal state in FeSi thin films demonstrates spin-momentum locking and nontrivial spin texture due to strong spin-orbit coupling and nonzero Zak phase of bulk band topology [7]. The related non equilibrium spin accumulation is shown to cause current-induced magnetization switching at moderate current densities ~$(1.6 \div 3.4)\cdot 10^7$ A/cm$^2$ [7-8]. This feature, which appears in $BaF_2$ capped FeSi 3 nm thin layers at room temperatures [8], opens a window for realization of strong spin-orbit coupling effects in the noble-metal-free silicon-integrated heterostructures in various spintronic applications. However, discovering the full potential of FeSi and other monosilicides as effective spintronic materials requires a detailed knowledge of electronic transport and magnetism that may be induced by non-zero Berry phase under various conditions (free-standing/capping surfaces, fabrication parameters, synthesis techniques etc.).

A powerful tool for detecting exciting properties related to nontrivial Berry phase is provided by Hall effect (HE) that appears in the related potential difference to be transverse to the applied current. Its rich potential is proved by numerous experimental studies that involve manifestation of intrinsic anomalous HE (AHE) originated from non-zero Berry phase [9], detection of topological HE induced by stable magnetic vortices (skyrmions) in MnSi single crystals and films [10-11] and discovering of hidden critical behavior in $Mn_{1-x}Fe_xSi$ initiated by frustrated indirect exchange interaction [12]. The HE family is complemented by planar HE (PHE), which is induced by in-plane applied magnetic field [13] and may originate from both chiral anomalies in Weyl semimetals [14-15] and/or lifted protection of surface Dirac states in topological insulators [16]. Angular resolved photoemission spectroscopy (ARPES) would discover the surface band structure of single crystals [17], but it fails to validate topological features in polycrystalline materials. Thus, magnetotransport technique is a unique strategy to detect chiral anomaly related to topological states in these alloys [14-15]. While non-vanishing

Berry phase in FeSi was predicted in 2013 [18], its experimental proof is missing so far. In this regard, the comparative analysis of transport phenomena in FeSi-based nanostructures can provide new insight about intrinsic magnetism and topological aspects of this prospective quantum material.

A notorious obstacle to such research is the challenge of achieving reliably controlled 1:1 stoichiometry in the Fe–Si system [19]. In this work, we overcome it by fabricating FeSi thin films via pulsed laser deposition (PLD) of $^{57}$Fe layer on native oxide-etched Si, followed by solid-phase reaction at 400 °C (Fig. 1a). To verify the quality of the polycrystalline ε-FeSi thin films, conversion electron Mössbauer spectroscopy (CEMS) [20,21] was applied that shows perfect stoichiometric composition of our FeSi thin films (Fig.1b). Other details of samples' preparation and characterization as well as those of experimental techniques are given in Supplemental Material (see also Figs. S1-S2)

**Temperature induced crossover in resistivity.** Zero field resistivity $\rho_{xx}$ of the FeSi thin films measured at temperatures 1.7 – 300 K is presented in Fig. 2a. At room temperatures $\rho_{xx}$(300K)=220 μΩ·cm agrees well with the data reported for bulk FeSi single crystals (200-240 μΩ·cm [22-24]) and epitaxial/polycrystalline FeSi thin films (140-280 μΩ·cm [25-26]) thus proving high quality of our ε-FeSi thin films. Rather high conductivity appears due to bulk charge carriers, which are thermally activated across the energy gap. The transport gap $\Delta_\rho \approx$ 36.4 meV estimated from the $ln(\rho_{xx}) = f(\Delta_\rho/2k_BT)$ graph agrees reasonably with 35.5±0.1 meV obtained previously for polycrystalline ε-FeSi films [23].

The difference between the $\rho_{xx}(T)$ data of our ε-FeSi thin films and bulk crystal is increased upon cooling (Fig.2a). Resistivity value $\rho_{xx} \approx$8.6 mΩ·cm detected at 1.7 K is lower than that of FeSi single crystal by four orders in magnitude (Fig.2a). This saturation of the resistivity could be related to surface conductivity established earlier for epitaxial ε-FeSi thin films [26]. However, below 15 K the $\rho_{xx}(T)$ data can be perfectly fitted by modified power dependence $\rho_{xx}^{LT}=1/(\rho_\infty^{-1}–A\cdot(T/T_1)^{-\delta})$ with $\rho_\infty \approx$4.77 mΩ·cm, A≈101.7 $(\Omega\cdot cm)^{-1}$ at $T_1$=1K and $\delta$=0.168≈1/6 (green dashed line in Fig.2a). Similar trend detected previously in niobium-carbon nanocomposite films with moderate disorder was ascribed to localization of charge carriers in weak random potential [27-28]. While the above mechanism is specific for 3D intergranular tunneling [28], any mixing effects between 3D and 2D conducting paths occurred in ε-FeSi thin films at low temperatures will be discussed elsewhere.

The data of Fig.2a point clearly to crossover from high-T conduction regime $\rho_{xx}^{HT}\sim\exp(–\Delta_\rho/2k_BT)$ controlled by thermally activated charge carriers to low-T trend $\rho_{xx}^{LT}(T)$. The combination of these contributions as two independent conducting channels $\rho_{xx}(T)^{-1}=(\rho_{xx}^{HT})^{-1}+(\rho_{xx}^{LT})^{-1}$ (grey double dotted line in Fig.2a) describes consistently the $\rho_{xx}(T)$ dependence in the full range of temperatures. Any deviations between calculated and measured data above 50 K (Fig. 2a) appear due to finite contribution of the low temperature component $\rho_{xx}^{LT}=\rho_{\infty}$ and may be easily corrected by increasing of $\Delta_{\rho}$ by ~3% (not shown here). At the same time, the ratio between ε-FeSi thin film and single crystal resistivity saturates to unity at room temperatures (red dash-dotted line in Figure 2a), proving that high-T electron transport above 100 K is controlled by conductivity of ε-FeSi crystallites rather than by currents through grain boundaries.

The temperature-induced crossover in resistivity becomes more pronounced if we introduce the factor of current inhomogeneity $CI=\rho_{xx}(T)^{-1}/((\rho_{xx}^{HT})^{-2}+(\rho_{xx}^{LT})^{-2})^{1/2}$, which equals to 1 when one conducting channel dominates and reaches $2^{1/2}$ if both channels contribute equally to conductivity. This non-monotonic parameter attains maximum under balanced low-T and high-T conductivities that became equal at 43 K (red line in Fig. 2b). Just the same behavior is found for the zero-field $R_{xy}/R_{xx}$ ratio estimated from experimental data (blue circles in Fig. 2b). Despite the very low values of $R_{xy}/R_{xx} < 0,3\%$ due to ideal arrangement of Hall probes, this experimental factor is in good agreement with the CI(T) dependence determined from the $\rho_{xx}(T)$ data. In our opinion, the fact that the zero-field transverse component of resistivity is not averaged out for these ε-FeSi films with polycrystalline texture opens up more possibilities for detection of effects related to Berry phase and non-trivial topology of electronic spectrum in quantum materials [9, 13].

**Magnetoresistance and Hall effect.** Temperature-induced crossover in electron transport is resolved in magnetoresistance (MR) data as well. The large positive MR achieving 12.8% at 14 T for 1.7 K (Figure S5a) decreases gradually with heating and changes its sign to the negative one above 20K. This negative MR component that reaches its minimum at 40K dominates up to 200 K when positive contribution from temperature excited charge carriers becomes comparable with the negative one (see the MR data for 275 K in Figure S5a). This behavior is particularly reminiscent of negative MR observed for bulk ε-FeSi crystals [23-24] and, together with negative longitudinal MR effect detected in [24], leaves room to search for chiral anomaly in FeSi thin films.

The temperature evolution and the sign of HE in ε-FeSi thin films (Fig. S5a) differ drastically from those ones known for FeSi single crystals. While only negative AHE in ε-FeSi single crystals was previously detected below 30 K [22-23], our ε-FeSi thin films exhibit positive AHE, the amplitude decreasing by more than three orders of magnitude with increasing temperature and remaining finite up to room temperature (Fig. 2d). On the contrary, the ordinary HE (OHE) component is field dependent and changes its slope to negative at temperatures over 20 K (Fig. 2c). Note that the gap $\Delta_R \approx 66$ meV estimated from activation law $R_H = \exp(\Delta_R/2k_BT)$ in the range of linear OHE (sell inset in Fig. 2c), is very close to $\Delta = 60$ meV for bulk FeSi single crystals [22-23]. The established correlation serves as additional proof of high quality of the FeSi thin films. At the same time, the noticeable discrepancy between $\Delta_\rho$ and $\Delta_R$ values seems to result from temperature dependent Hall mobility $\mu_H = R_H/\rho$ (not shown here), which follows the power law asymptotic $\mu_H \sim T^\eta$, $\eta \approx -3/2$ between 60K and 150K and is specific for high temperature electron transport in bulk FeSi crystals [22-23].

Resuming the AHE behavior in the ε-FeSi note that the coercive field of AHE loop falls down from 0.1 T at 1.7 K to 2 mT at 250 K (open circles in Fig. 2d). This evolution is fundamentally different from that of coercive force found from magnetization (M) measurements, which decreases from $H_C^M \approx 18$ mT at 1.7 K to $H_C^M \approx 8$ mT at 400K and can be attributed to tiny Fe surface imperfections (see Experimental Samples in Supplemental Material). The quantitative distinction between HE and M loops (Figure S4a) proves definitely that the observed AHE is not related to any spontaneous magnetic moment of the investigated ε-FeSi thin film and should have intrinsic origin to be different from conventional ferromagnets.

**Universal scaling of AHE in FeSi.** Temperature crossover in electron transport in FeSi (Fig.2a-d, see also Fig.S5) can be intuitively linked to transition from high-T 3D behavior (T>80 K), which is characteristic of bulk narrow gap semiconductor [22-24], to low-T 2D conductivity previously observed in the surface metallic layer in epitaxial films [7-8]. This surface conductivity was confirmed experimentally by detailed comparison of low-T conductivity and Hall effect measured for epitaxial FeSi 5-60 nm thin films capped with 10-nm-thick MgO [7]. Leaving these low-T anomalies, which are rather different from those of FeSi single crystals (positive sign of AHE, positive MR below 12 K, low values of saturated resistivity etc. [22-24]), we focus below on the specific features of topological electron transport in this non-centrosymmetric binary compound.

Basically, AHE is contributed from intrinsic and extrinsic mechanisms, which are related to the Berry curvature integrated over Brillouin zone and to impurity scattering, respectively [9]. These two regimes may change each other under external factors, e.g., temperature damping of spin fluctuations in related MnSi compound results in AHE crossover from extrinsic $\rho_{xy}^{AHE} \sim \rho_{xx}$ behavior detected in paramagnetic phase to intrinsic $\rho_{xy}^{AHE} \sim \rho_{xx}^2$ trend found below Curie point [30]. In our ε-FeSi thin films a universal scaling of AHE contribution by square of resistivity $\rho_{xy}^{AHE}$~$\rho_{xx}^2$ is perfectly established for temperatures between 1.7 K and 225 K (circles in Fig. 3c).

To verify independently this intriguing finding, we have also compared the $\rho_{xy}^{AHE}$ and $\rho_{xx}$ data reported earlier for bulk ε-FeSi single crystals [22-23, 31], crystalline MBE ε-FeSi films [7-8] and off-stoichiometric amorphous $Fe_xSi_{1-x}$ films (0.43<x<0.71) [32]. While all the $\rho_{xy}^{AHE}(\rho_{xx})$ data expressed in bulk units (Ω·cm) show rather different behavior for thin films and single crystals (Fig.3a), all the data except for the off-stoichiometric compositions (x≥0.55) merge almost perfectly into one universal $\rho_{xy}^{AHE}(\rho_{xx})$ dependence if reduce them to surface units (Ω per square, Fig.3b). In the last case the general trend is very similar to $\rho_{xy}^{AHE}$~$\rho_{xx}^2$ law shown by grey dashed line in Fig.3b. Note here that this universal $\rho_{xy}^{AHE}(\rho_{xx})$ behaviour is not disturbed by either temperature crossover in conductivity (Fig. 2a-b) or the polycrystalline texture of ε-FeSi thin films consisting of 40 nm sized crystallites (Fig. 1a). This result convincingly demonstrates that AHE in ε-FeSi is intrinsic in nature and arises from non-zero Berry curvature. In our opinion, almost perfect correlation between our $\rho_{xy}^{AHE}(\rho_{xx})$ data (open triangles in Fig.3b) and these ones reported for single crystals and MBE films (datasets 2-3 and 5-8 in Fig.3b, respectively), proves high quality of the polycrystalline films and provides opportunity to study the specific features induced by the non-trivial topology.

**Chiral anomaly in ε-FeSi thin films.** Another specific feature arising from non-trivial topology of solids is a chiral anomaly, which is generally attributed to the topological state of Weyl semimetal [13, 29]. Electron transport in this topological material is contributed by quasiparticles, which act like massless relativistic Weyl fermions near Weyl points and are essentially "bulk phenomena" [29]. Two apparent manifestations of chiral anomaly in Weyl semimetal are anisotropic negative longitudinal magnetoresistance (anisotropic magnetoresistance, AMR), which appears due to suppressed scattering of electrons of opposite chirality in crossed in-plane magnetic and electric fields [29], and planar Hall effect (PHE), which emerges in a transverse voltage under not fully aligned coplanar electric and magnetic fields [13]. The common origin of AMR and PHE fixes for them the same amplitude to be equal to the difference $\Delta\rho_{chiral}=R_{\|}-R_{\perp}$ between the resistances parallel with and perpendicular to in-plane magnetic field, respectively, but with π/4 phase difference for corresponding oscillations (~$\cos^2(\varphi)$ for AMR and ~$\sin(\varphi)\cdot\cos(\varphi)$ for PHE [13]). Principally, the AMR and PHE amplitudes depend on chiral charge imbalance, which is proportional to both applied electric and magnetic fields. The above leads to the conclusion that the chiral anomaly should be initially looked for in the temperature range where bulk electron transport and negative magnetoresistance are observed simultaneously. This corresponds exactly to the right slope of $R_{xy}/R_{xx}$ at T>40 K (Fig. 2b) where bulk conductivity is comparable to the surface one or predominates in total conductivity of the ε-FeSi thin films.

Addressing to PHE, this phenomenon is induced by anisotropy of magnetoresistance with respect to in-plane magnetic field applied along and perpendicular to electric field [13]. At all the temperatures reported below, we found negative longitudinal magnetoresistance with well-resolved anisotropic behavior (Fig.4a). The amplitude was found to oscillate with φ variation as $\rho_{xx}^{AMR} \sim \cos^2(\varphi)$ (Fig.4b). The extremely low amplitude of AMR oscillations does not exceed 0.1 % of total resistivity (Figs.3a-b). For all measurements the phase shift $\varphi_0$ determined from free parameter nonlinear fitting by $\rho_{xx}^{AMR}\sim\cos^2(\varphi-\varphi_0)$ did not exceed $1^0$.

The occurrence of AMR in the ε-FeSi thin films proves search for PHE. The $\rho_{xy}(B;\varphi,\theta=0^0)\equiv\rho_{xy}(B;\varphi)$ data presented in Fig.4c show that both shape and amplitude of the $\rho_{xy}(B;\varphi)$ loop vary significantly under in-plane rotation of magnetic field. This complicated behavior results from several transverse effects induced by in-plane rotation of applied magnetic field ($\theta=0^0$). The amplitude of the odd contribution, which is reminiscent that of AHE (noted below as *longitudinal Hall effect, LHE*) can be estimated in zero magnetic field as the averaged difference $\rho_{xy}^{LHE}(\varphi_0)=(\rho_{xy}(+0T;\varphi_0)-\rho_{xy}(-0T;\varphi_0))/2$. The marks of +0T and –0T correspond to the residual $\rho_{xy}^{LHE}$ values after sweeping out of magnetic field from maximal positive and negative values (Fig.4c). To extract PHE from field dependent $\rho_{xy}(B;\varphi)$ data (Fig.4c) we averaged the data taken at opposite directions of applied magnetic field: $\rho_{xy}^{PHE}(\varphi)=(\rho_{xy}(+B;\varphi)+\rho_{xy}(-B;\varphi))/2$ [13]. Here and below we refer to the data at a fixed magnetic field of B=0.7 T, the field dependence of PHE will be discussed elsewhere.

The extracted dependencies of $\rho_{xy}^{PHE}(\varphi)$ and $\rho_{xy}^{LHE}(\varphi)$ (Fig.4d) demonstrate oscillations with different frequencies. While $\rho_{xy}^{PHE}(\varphi)$ follows the predicted dependence of $\rho_{xy}^{PHE}(\varphi)=\rho_{xy}^{PHE}\cos(\varphi)\sin(\varphi)$, $\rho_{xy}^{LHE}(\varphi)$ exibits the cos(φ) behavior, which is very similar to the oscillation of OHE under transverse magnetic field rotation (θ variation under $\varphi=90^0$, not shown here). Despite similar temperature trends in the $\rho_{xy}^{PHE}(T)$ and $\rho_{xy}^{LHE}(T)$ amplitudes, one can note a clear distinction when $\rho_{xy}^{PHE}$ exceeds $\rho_{xy}^{LHE}$ by 30% at 43K and, vice versa, $\rho_{xy}^{LHE}$ exceeds $\rho_{xy}^{PHE}$ by 34% at 125K (the inset in Fig.4d). This diversity confirms different nature of PHE and LHE in the ε-FeSi thin films. Moreover, the correlated oscillations in $\rho_{xy}^{PHE}$ (Fig.4d) and $\rho_{xx}^{AMR}$ (Fig.4b) with the equal amplitudes ($\rho_{xy}^{PHE}=\rho_{xx}^{AMR}=\Delta\rho_{chiral}=(6.1\pm0.5)$ μΩ·cm at 43 K) and π/4 phase difference allow us to assign unambiguously the observed PHE and AMR to chiral anomaly that originates from the Weyl semimetal state inherent to the bulk electronic spectrum of ε-FeSi.

**Discussion.** Universal scaling of anomalous Hall resistivity in ε-FeSi discovered for the first time in this work (Fig.3b), allows to compare scaling dependences for AHE, LHE and PHE components measured in the polycrystalline films (Fig.3c). Rather nice scaling $\rho_{xy}^{LHE}$~$\rho_{xx}^2$ is also observed for the LHE component (open triangles in Fig.3c). Hypothetically, this LHE behavior may be related to that of AHE if suggest that LHE arises exclusively due to some misorientation of FeSi film (~$5^0$ if taken $\sin(\theta_0)=\rho_{xy}^{LHE}/\rho_{xy}^{AHE}$). However, the estimated misalignment is quite large to be occasional and does not explain the difference established in the temperature dependences of the AHE/LHE amplitudes and coercive forces (Figure S4b).

The difference between AHE and LHE becomes more visible when reduce the resistivity data of Fig.3c to the components of conductivity tensor: $\sigma_{xx}=\rho_{xx}\cdot d/(\rho_{xx}+\rho_{xy})^2$, $\sigma_{yx}=\rho_{xy}\cdot d/(\rho_{xx}+\rho_{xy})^2$. The film thickness *d* was added to scale all these data to the unit film area not considering any crossover between high-T 3D and low-T 2D regimes of conductivity. Temperature independent behavior of anomalous Hall conductivity $\sigma_{xy}^{AHE}$~const ($\sigma_{xx}$) staying almost constant at $\sigma_{xy}^{AHE}\approx 14$ μS/sq. below 200 K (Fig.3d) corresponds to intrinsic AHE in this compound. At the same time, below 80K $\sigma_{xy}^{LHE}$ depends on excitation current falling down by half at 1 mA and increasing slightly at 5 mA (Fig. 3d). This is not the case of $\sigma_{xy}^{AHE}$, which does not depend on excitation current (lower $\sigma_{xy}^{AHE}$ values for 1mA in Fig. 3d were detected in another experimental run and present an upper bound of the experimental error). The different trends in $\sigma_{xy}^{AHE}(\sigma_{xx})$ and $\sigma_{xy}^{LHE}(\sigma_{xx})$ point definitely to different origins of AHE and LHE in ε-FeSi thin films. In turn, current dependent behavior of $\sigma_{xy}^{LHE}(\sigma_{xx})$ below 80K ($\sigma_{xx}<5$ mS/sq., Fig.3d) may be induced by drastic increase of effective current density (from $2.6\cdot 10^3$ A/cm$^2$ up to $4\cdot 10^5$ A/cm$^2$ if suggest single atomic conducting layer) when entering the low-T conductivity regime (Figs.2a-b). However, even in this case the current density stays well below the critical level of $1.6\cdot 10^7$ A/cm$^2$ that induces residual magnetization switching in FeSi films [7].

Extending the discussion, we note that universal scaling of AHE (Fig.3) has nothing to do with the firstly discovered PHE, which demonstrates an exotic power-law evolution between 40K and 80 K described by $\rho_{xy}^{PHE}$~$\rho_{xx}^{2\alpha}$ or, equivalently, to $\rho_{xy}^{PHE}$~$(\rho_{xy}^{AHE})^{\alpha}$ ($\alpha$=4/3) (solid squares in Fig.3c). This parametric dependence becomes more pronounced in units of conductivity where amplitude of PHE follows to power law $\sigma_{xy}^{PHE} \sim \sigma_{xx}^{\alpha}$ ($\alpha = -2/3$) below 80K (Fig.3d). We have not found any similar precedents of scaling behavior in other PHE studies [29]. Thus, a further extensive study of the suggested chiral anomaly in anisotropic longitudinal magnetoresistance and PHE is required to shed light on the origin of exotic PHE scaling behavior discovered in this work.

Returning to the observed temperature-independent scaling of anomalous Hall conductivity $\sigma_{xy}^{AHE}$~const ($\sigma_{xx}$) (with a nearly constant value $\sigma_{xy}^{AHE}\approx 14$ μS/sq. ), it is worth emphasizing that this scaling behavior appears to be decoupled from the bulk magnetization. Indeed, our systematic analysis revealed no discernible correlation between the magnetization and any of the AHE parameters. This finding allows suggesting that this anomalous Hall conductivity appears due to Berry phase contribution from Weyl points in the ε-FeSi band structure rather than due to some features of spin polarized electronic structure. In this regard it is worth evaluating the distance between the momentum locations of two Weyl points $k_+^W$ and $k_-^W$ (Fig.1e) if one applies the general relation for Hall conductivity $\sigma_{xy}=(e^2/h)(k_+^W-k_-^W)/(2\pi)$ [25]. Straightforward calculation results in $(k_+^W-k_-^W)/(2\pi)\approx 0{,}36$ that corresponds to momentum location $k_+^W=-k_-^W\approx\pi/3$. The reasonable value requires careful revisiting of low-energy electronic spectrum of FeSi and related B20 materials to reveal factors, which allow the Fermi energy to approach the Weyl nodes under controlled conditions (temperature, composition, pressure etc.).

**Conclusion.** Based on comprehensive Hall effect study of high-quality polycrystalline ε-FeSi thin films, we have firstly revealed the universal $\rho_{xy}^{AHE}\sim\rho_{xx}^2$ scaling of anomalous Hall resistivity to be valid for both bulk crystals and thin films when reducing $\rho_{xy}$ and $\rho_{xx}$ to surface units. The related temperature independent anomalous Hall conductivity $\sigma_{xy}^{AHE}$~const($\sigma_{xx}$) unambiguously indicates the intrinsic nature of AHE originated from non-zero Berry curvature [9]. To the best of our knowledge, this is the first direct experimental demonstration of non-vanishing Berry phase in the chiral insulator FeSi predicted more than ten years ago [18]. Our conclusion is supported by observation of chiral anomaly with correlated AMR and PHE behavior $\sigma_{xy}^{PHE}\sim(\sigma_{xx})^{\alpha}$ ($\alpha= -2/3$) below 80 K, which we attribute to the specific feature of chiral anomaly in these polycrystalline ε-FeSi thin films.

The detection of chiral anomaly in AMR and PHE in the presence of crystalline disorder provides strong evidence that ε-FeSi is a new high temperature Weyl semimetal compound. The important advantage of FeSi is that both Fe and Si are naturally abundant and do not contain potentially critical rare-earth elements. This fact, along with the proposed recipe to synthesize polycrystalline FeSi directly integrated on Si-substrates, is enormously relevant from a technology-transfer perspective, opening a path toward new CMOS compatible options to exploit nontrivial topology in condensed matter.

**Acknowledgements.** The authors thank Shared Facility Centre of MIPT, Mendeleev Center for Collective Use of Mendeleev University of Chemical Technology and Shared Facility Centre for Studies of HTS and other Strongly Correlated Materials of Lebedev Physical Institute. The study of electron transport in the FeSi thin films (A. Bozhko, V. Zhurkin, A. Bogach and V. Glushkov) was supported by a grant from the Russian Science Foundation No. 25-72-20032 (https://rscf.ru/en/project/25-72-20032/). R. Mantovan and A. Zenkevich acknowledge the experiment IS-578 conducted at ISOLDE/CERN and the EU H2020 project ENSAR2 (Grant no. 654002) for having initiated the idea of conducting the experiments here presented.

**Authorship contribution statement. R. Mantovan:** Original idea, Writing – review & editing, Writing – original draft, Validation, Methodology, Investigation, Formal analysis, Project administration, Data curation, Conceptualization. **A. Bozhko:** Writing – original draft, Validation, Methodology, Investigation, Formal analysis, Data curation, Conceptualization. **V. Zhurkin:** Validation, Methodology, Investigation, Formal analysis. **A. Bogach:** Validation, Methodology, Investigation. **A. Khanas:** Writing – original draft, Validation, Methodology, Investigation, Formal analysis. **S. Zarubin:** Validation, Methodology, Investigation. **A. Zenkevich:** Writing – review & editing, Writing – original draft, Validation, Supervision, Methodology, Formal analysis. **V. Glushkov:** Theoretical analysis, Writing – review & editing, Writing – original draft, Validation, Supervision, Project administration, Methodology, Investigation, Funding acquisition, Formal analysis, Conceptualization.

**Declaration of competing interest.** The authors declare that they have no known competing financial interests or personal relationships that could have appeared to influence the work reported in this paper.

**References.**

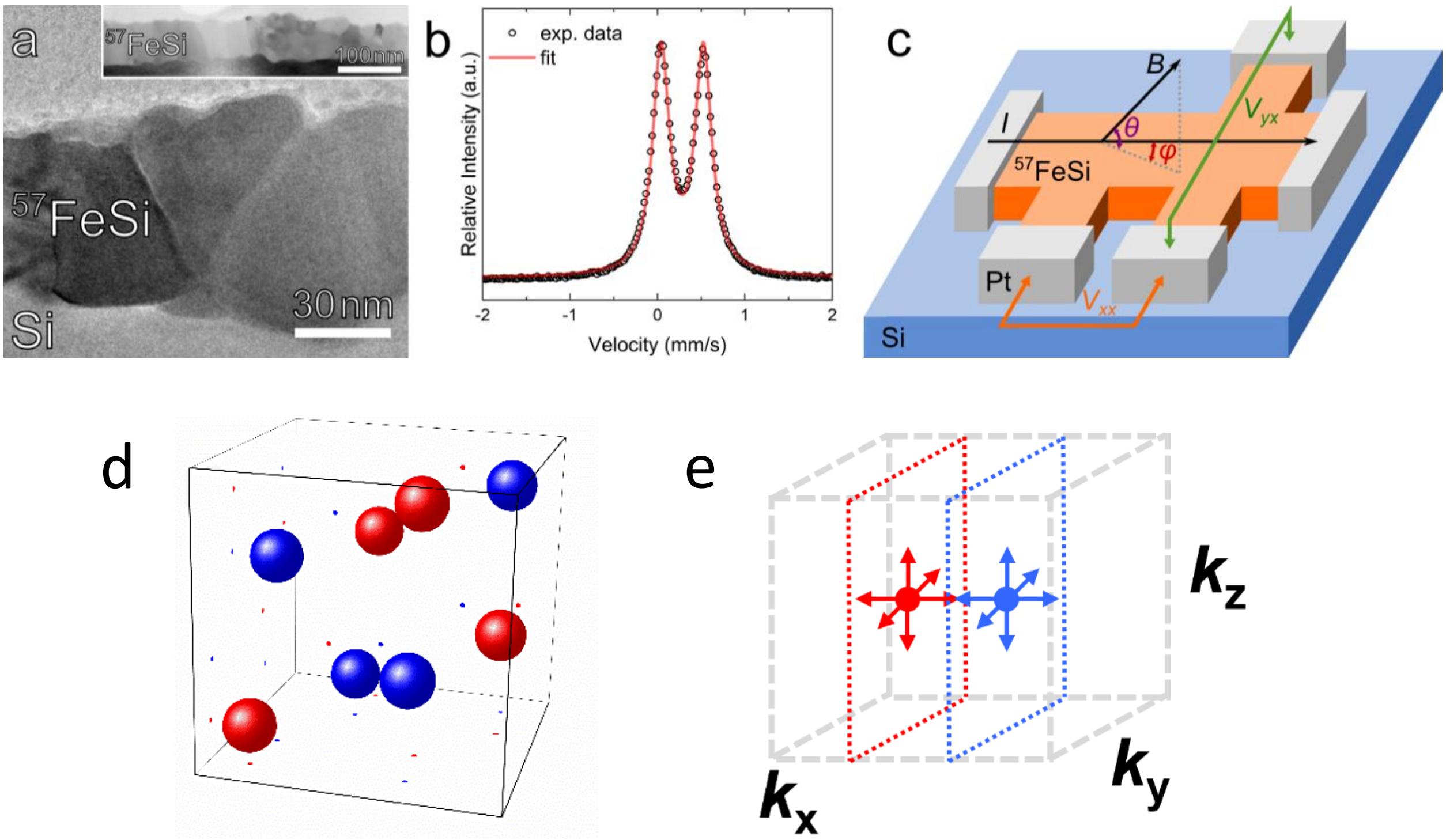

**Figure 1.** (a) TEM images of the PLD grown ε-$^{57}$FeSi thin films. (b) CEMS spectrum of polycrystalline ε-$^{57}$FeSi thin film. (c) Experimental scheme of the fabricated sample used to measure longitudinal ($V_{xx}$) and transverse ($V_{xy}$) voltages under different directions of magnetic field B (fixed by azimuthal φ and polar θ angles with respect to applied excitation current). (d) Unit cell of B20 ε-FeSi with 4 formulae units, each atom being projected on the *xy*-, *yz*- and *xz*-planes. (e) Sketch of two Weyl points separated in *k*-space of Brillouin zone along *x* axis. Similar pairs of Weyl points along y and z axes are not shown for clarity.

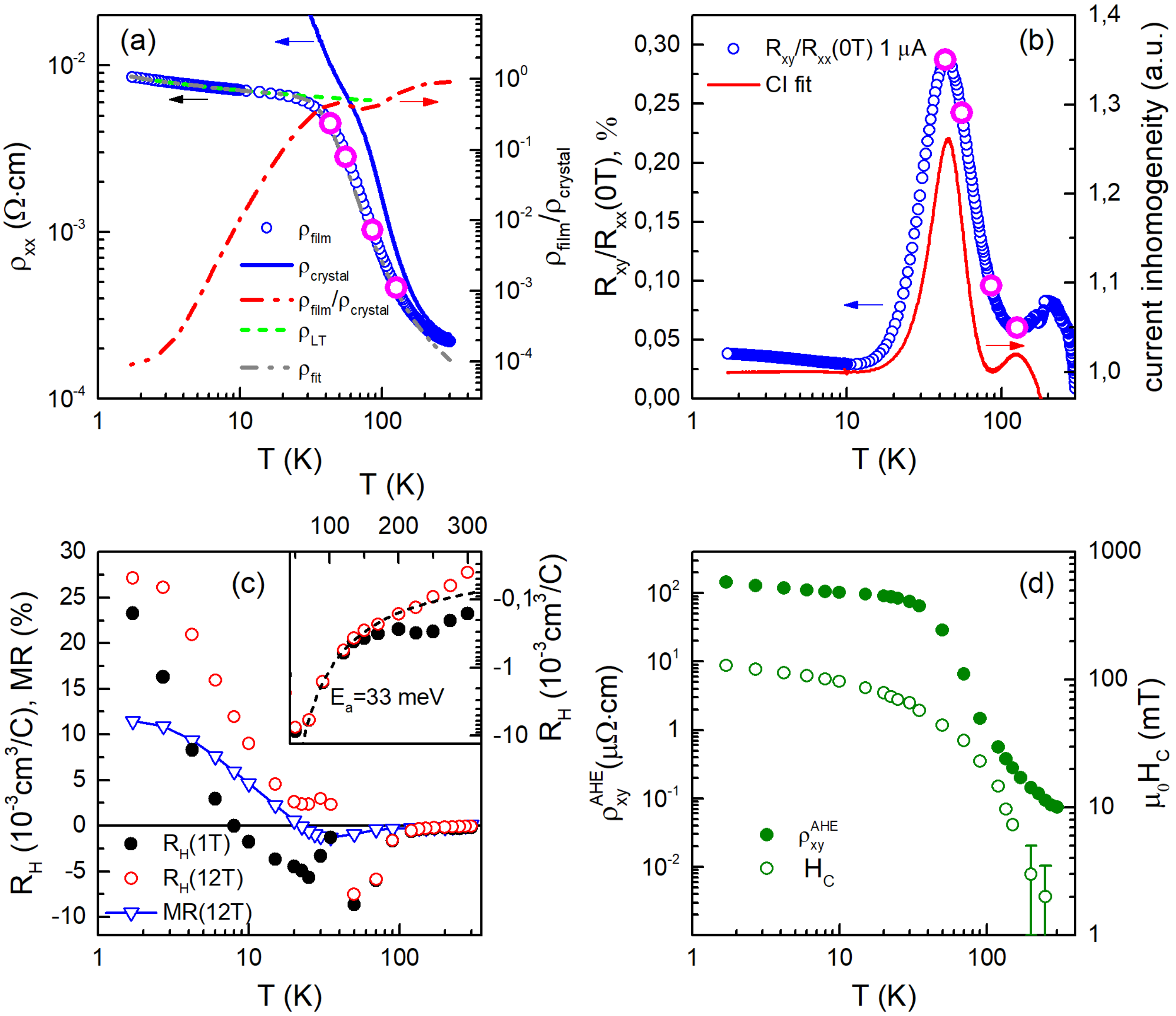

**Figure 2.** (a) Temperature dependences of $\rho_{xx}$ resistivity measured for 65 nm ε-FeSi thin film ($\rho_{film}$) at different excitation currents (noted in the plot). Blue solid and red dash-dotted lines show the reference data for FeSi single crystal ($\rho_{crystal}$) and the relation $\rho_{film}/\rho_{crystal}$, respectively. Green and grey curves correspond to low temperature contribution to resistivity ($\rho_{LT}$, green short-dash) and general two-component fit ($\rho_{fit,}$ grey double dotted) discussed in the text.

(b) Zero-field transverse to longitudinal resistance $R_{xy}/R_{xx}$ ratio (left axis) compared to current inhomogeneity factor CI(T) (right axis). Open circles in panels a and b correspond to temperatures of PHE measurements.

(c) Low field (1T) and high field (12T) Hall coefficient $R_H$ compared to high field magnetoresistance amplitude MR=Δρ/ρ extracted from raw experimental data (Fig.S5, [17b]). Inset presents high temperature zoom with activation fit shown by red dotted line.

(d) The amplitude of anomalous Hall resistivity $\rho_{xy}^{AHE}$ measured in transverse (Θ=$90^0$) magnetic field (closed symbols, left y axis) compared to the corresponding magnetic coercive force $\mu_0H_C$ (open symbols, right y axis).

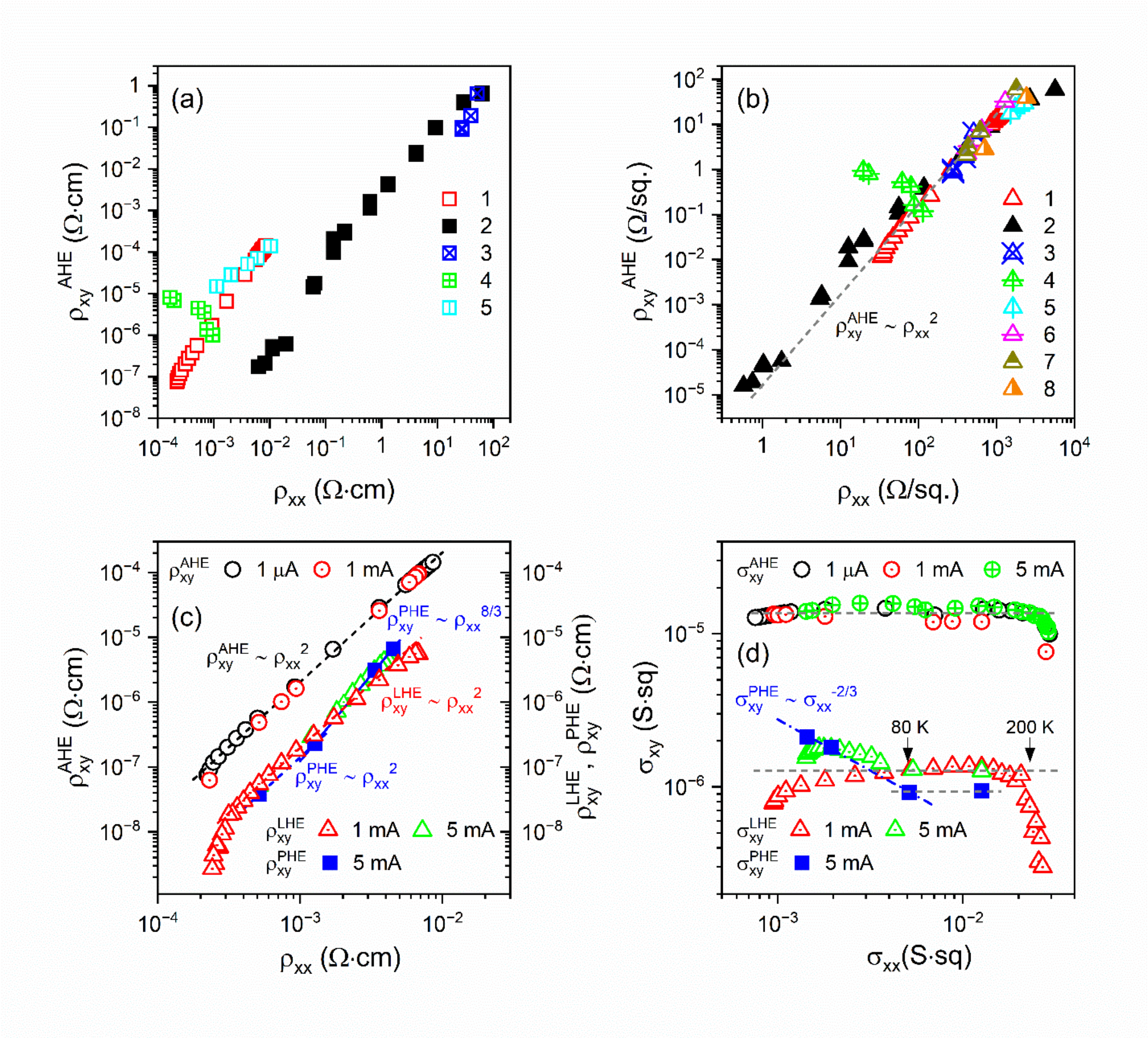

**Figure 3.** Scaling of anomalous Hall resistivity $\rho_{xy}^{AHE}$ vs $\rho_{xx}$ calculated in (a) bulk and (b) surface units for polycrystalline (PC) ε-FeSi films (1), ε-FeSi single crystals [23] (2), [22] (3), amorphous $Fe_xSi_{1-x}$ [32] (4), and crystalline MBE ε-FeSi films: MgO capped [7] (5), $CaF_2$ capped [8] (6), $BaF_2$ capped [8] (7), MgO capped [8] (8). All the details (thicknesses and temperature ranges) are given in Supplementary Materials.

(c) Double-log plot of anomalous ($\rho_{xy}^{AHE}$, circles), longitudinal ($\rho_{xy}^{LHE}$, triangles) and planar ($\rho_{xy}^{PHE}$, squares) Hall resistivity as a function of $\rho_{xx}$ in PC ε-FeSi films calculated in bulk units. Grey dashed lines in panels (b-c) show $\rho_{xy}^{AHE}$~$\rho_{xx}^2$ fits that correspond to intrinsic anomalous Hall effect due to non-zero Berry curvature. Red and blue dotted lines are related to squared resistivity scaling of longitudinal and planar Hall resistivities $\rho_{xy}^{LHE}$,$\rho_{xy}^{PHE}$~$\rho_{xx}^2$. Blue solid line is the best fit for $\rho_{xy}^{PHE}$~$\rho_{xx}^{\alpha}$ with exponent $\alpha$=8/3. The scales of left and right y axes are identical.

(d) The $\rho_{xy}^{AHE}$, $\rho_{xy}^{LHE}$and $\rho_{xy}^{PHE}$ data recalculated in terms of conductivity $\sigma_{xy}^{AHE}$, $\sigma_{xy}^{LHE}$and $\sigma_{xy}^{PHE}$ (in Siemens per square) as a function of $\sigma_{xx}$. Horizontal grey dashed lines mark behaviour corresponded to intrinsic anomalous Hall effect due to non-zero Berry phase.

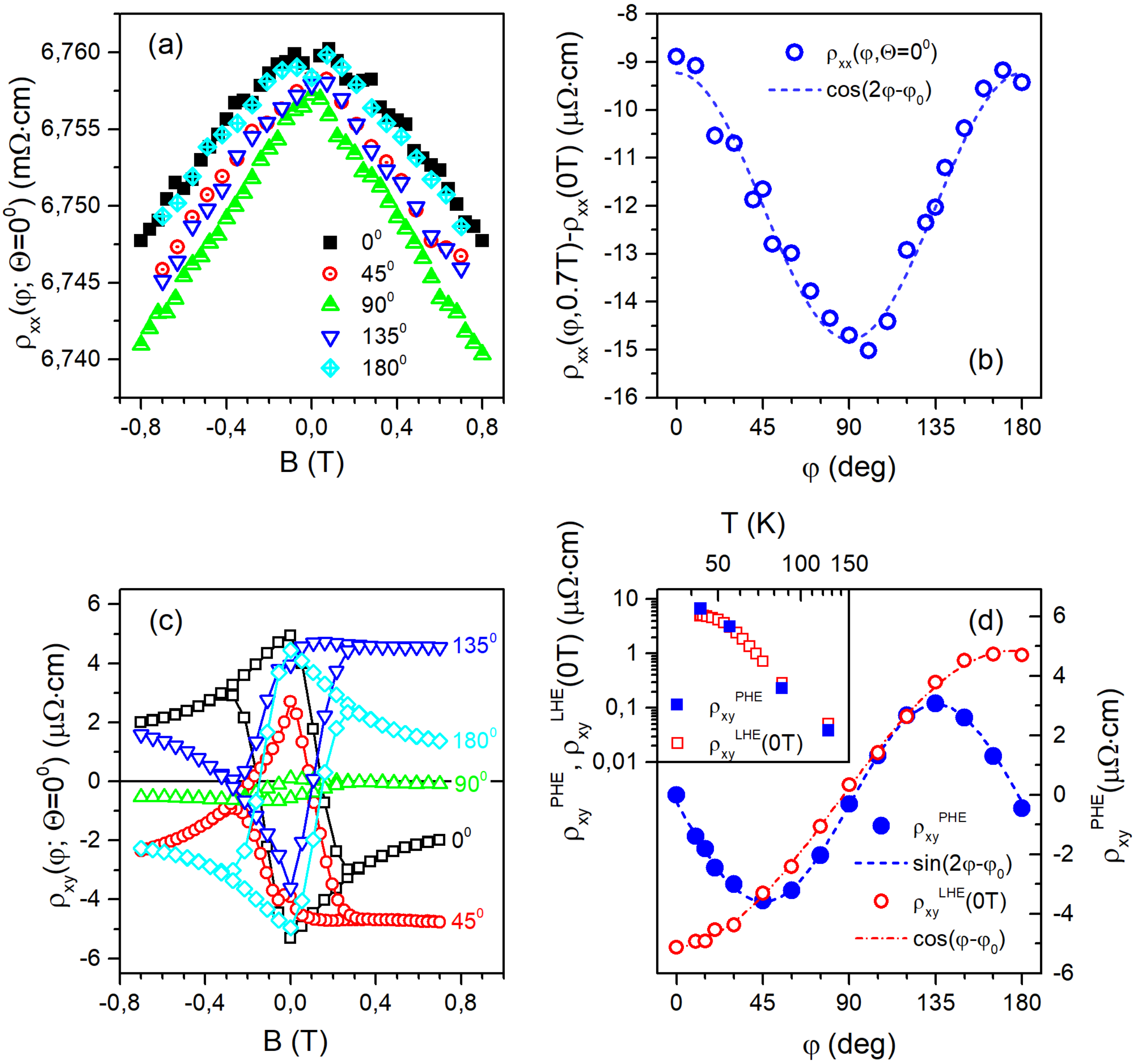

**Figure 4.** (a) Low-field anisotropy of in-plane resistivity $\rho_{xx}(B, \varphi; \Theta=0^0)$ and (c) low-field hysteresis loops in planar Hall resistivity $\rho_{xy}(B, \varphi; \Theta=0^0)$ measured for 65 nm FeSi thin film at 43 K by using 5 mA excitation current. The numbers in the legends mark azimuthal angle $\varphi$. (b,d) Fitting of (b) anisotropic magnetoresistance data and (d) planar and zero-field longitudinal Hall resistivity by using of $\rho_{xx}^{AMR}(\varphi)=\rho_{xx}(\varphi,B=0.7\ T)-\rho_{xx}(0\ T)=\rho_{xx}^{AMR}\cdot\cos^2(\varphi-\varphi_0)$ ($\rho_{xx}(0\ T)=6.7583$ mΩ·cm) and $\rho_{xy}^{PHE}(\varphi)=\rho_{xy}^{PHE}\cdot\sin(\varphi-\varphi_0)\cdot\cos(\varphi-\varphi_0)$, $\rho_{xy}^{LHE}(\varphi)=\rho_{xy}^{LHE}\cdot\cos(\varphi-\varphi_0)$ , respectively. The $\rho_{xy}^{PHE}(\varphi)$ data are calculated from the data of Fig.4c as the average taken at +/–0.7T. Inset in panel (d) shows the temperature dependences of the $\rho_{xy}^{PHE}$ and $\rho_{xy}^{LHE}$ amplitudes.

Supplemental material to

R. Mantovan, A. Bozhko, V. Zhurkin, A. Bogach, A. Khanas, S. Zarubin, A. Zenkevich, V. Glushkov.

**“Experimental Signatures of Topological Transport in Weyl semimetal candidate FeSi ”**

## Experimental samples

A 30 nm thick Fe layer enriched at 95% with $^{57}Fe$ isotope to enable conversion electron Mössbauer spectroscopy (CEMS) was deposited by Pulsed Laser Deposition (PLD) technique on the Si (100) substrate with natural $SiO_2$ layer chemically etched prior to deposition. FeSi thin film 65 nm in thickness was formed by a solid-state reaction following *in situ* vacuum thermal annealing at T = 400 °C for 4 hours (Fig. S1(a)-(c)). The formation of ε-FeSi phase with the cubic B20 crystal structure was confirmed by electron diffraction (Fig. S2) and X-ray diffraction (Fig. S3(a)). The thickness of FeSi layer was estimated to be ~65 nm from TEM images. Exhaustive analysis of electron diffraction patterns has allowed to rule out the presence of any inclusions with different symmetry beside B20 FeSi (Fig. S2(b)), such as α-Fe (diffraction circles of α-Fe do not correspond to any reflexes seen in Fig. S2(c)).

The local environment of Fe atoms in the grown FeSi layer was also analysed by CEMS that is unique in identifying and quantifying all the Fe-related phases even if not crystalline. The CEMS data confirmed that the films synthesized are constituted by ε-FeSi with hyperfine parameters very well matching those expected for the pure and stoichiometric B20 phase (Fig. S3(b)). Only tiny features, corresponding to a residual ~1 % Fe fraction, were detected in XRD and CEMS (Figs. S3(a, b)), which are attributed to Fe droplets originating from laser ablation and sparsely spread over the surface of the sample. Direct comparison of Hall effect (HE) and magnetization hysteresis curves (Fig. S4a) reveal different values and temperature behavior of the related parameters (residual magnetization and coercive force, Fig. S4b) that allows us to exclude any contribution to HE features from the free Fe droplets located on the surface of ε-FeSi films.

As-grown FeSi thin films were patterned by photolithography and plasma etching into the Hall bar geometry to allow the measurement of both longitudinal ($V_{xx}$) and transverse ($V_{yx}$) voltages, Fig. S1(d-e).

## Experimental methods

Electron transport measurements were performed with the help of Quantum Design PPMS setup (*ac* current) down to 1.7K in magnetic fields up to 14 T. Resistivity measured in traditional scheme was calculated as $\rho_{xx}=R(b\cdot d/l)$, where R – the resistance between one-side potential probes. We used the only configuration of Hall bar device with the distance between the centers of potential probes l=2.93 mm and with the width and thickness of the film, b=3 mm and d=65 nm, respectively (Fig. S1(f)). Hall resistivity was calculated as $\rho_{xy}(T)=R_{xy}\cdot d$, $R_{xy}$ – the resistance between transverse potential probes averaged for two opposite directions of magnetic field applied perpendicular the film’s plane ($\Theta=90^0$) and d=65 nm – the thickness of the film. Raw data of magnetoresistance ($\Delta\rho/\rho$) (b) and Hall resistivity ($\rho_{xy}$) are shown in Fig. S5.

To reveal temperature evolution of anomalous (AHE) and ordinary (OHE) Hall effects, the experimental isotherms $\rho_{xy}(B,T_0)$ (Fig. S5b) were fitted by linear trends $\rho_{xy}(B)=\rho_{xy}^{AHE}+R_H\cdot B$ ($R_H$ – Hall coefficient) in low magnetic fields 0.4÷2 T. The $R_H$(1T) data compared to high-field Hall coefficient $R_H(12T)=(\rho_{xy}(12T)-\rho_{xy}^{AHE})/12T$ are shown by closed and open circles, respectively, in Figure 2c of the main text.

The longitudinal HE (LHE) and planar HE (PHE) data in Figure 3 were conducted in a closed-cycle a ARS cryostat under external (variable angle) magnetic fields up to 0.8 T. Constant current were supplied through a Keithley 6221 current source and voltages recorded with a Keithley 2182A nanovoltmeter. Samples were mounted in a custom-made sample holder to accommodate and electrically contact the Hall bar device. To extract LHE and PHE the hysteretic curves of $R_{xy}(B,\varphi,\Theta=0^0)$ were measured with magnetic field applied in the plane of the film ($\varphi=0^0$ for LHE). Optionally, van der Pauw method [S1] was applied to check both the films' homogeneity and a reproducibility of the $\rho_{xx}$ and $\rho_{xy}$ data obtained under mutual exchange of current and potential probes.

Magnetic moment of the films was measured by Quantum Design MPMS-5-XL SQUID magnetometer at temperatures 1.8-400 K in magnetic fields up to 5 T.

**Anomalous Hall resistivity**

To verify a universal scaling of AHE contribution by square of resistivity $\rho_{xy}^{AHE}\sim\rho_{xx}^2$ we have also compared our $\rho_{xy}^{AHE}$ and $\rho_{xx}$ data with those ones reported earlier for bulk $\varepsilon$-FeSi single crystals [22-23, 31], crystalline $\varepsilon$-FeSi films grown by molecular beam epitaxy (MBE) [7-8] and off-stoichiometric amorphous $Fe_xSi_{1-x}$ films (0.43<x<0.71) [32]. The effective parameters (thickness and temperature range) of the investigated samples (in notations of Figs.3a-b and references from the main text) are listed below:

1) polycrystalline $\varepsilon$-FeSi films: d=65 nm, T=1.8-300K (this work);
2) $\varepsilon$-FeSi single crystals: d=0,12 mm, T=1.8-80K [23,31];
3) $\varepsilon$-FeSi single crystals: d=1 mm, T=1.58-3.6K [22];
4) amorphous films of $Fe_xSi_{1-x}$ (0.43<x<0.71): d= 50-100 nm, T=2K [32];
5) MgO capped crystalline MBE $\varepsilon$-FeSi films: d= 5-60 nm, T=2K [7];
6) $CaF_2$ capped crystalline MBE $\varepsilon$-FeSi films: d= 10 nm, T=2-300K K [8];
7) $BaF_2$ capped crystalline MBE $\varepsilon$-FeSi films: d= 10 nm, T=2-300K K [8];
8) MgO capped crystalline MBE $\varepsilon$-FeSi films: d= 10 nm, T=2-100K K [8].

The data (6)-(8), which were calculated from the $\sigma_{xy}^{AHE}$ and $\sigma_{xx}$ data extracted in [8] in surface units, are not presented in Fig.3a. The thickness of $\varepsilon$-FeSi single crystals not mentioned by the authors in [22] was taken at the reasonable value d=1 mm not influencing on the overall scaling.

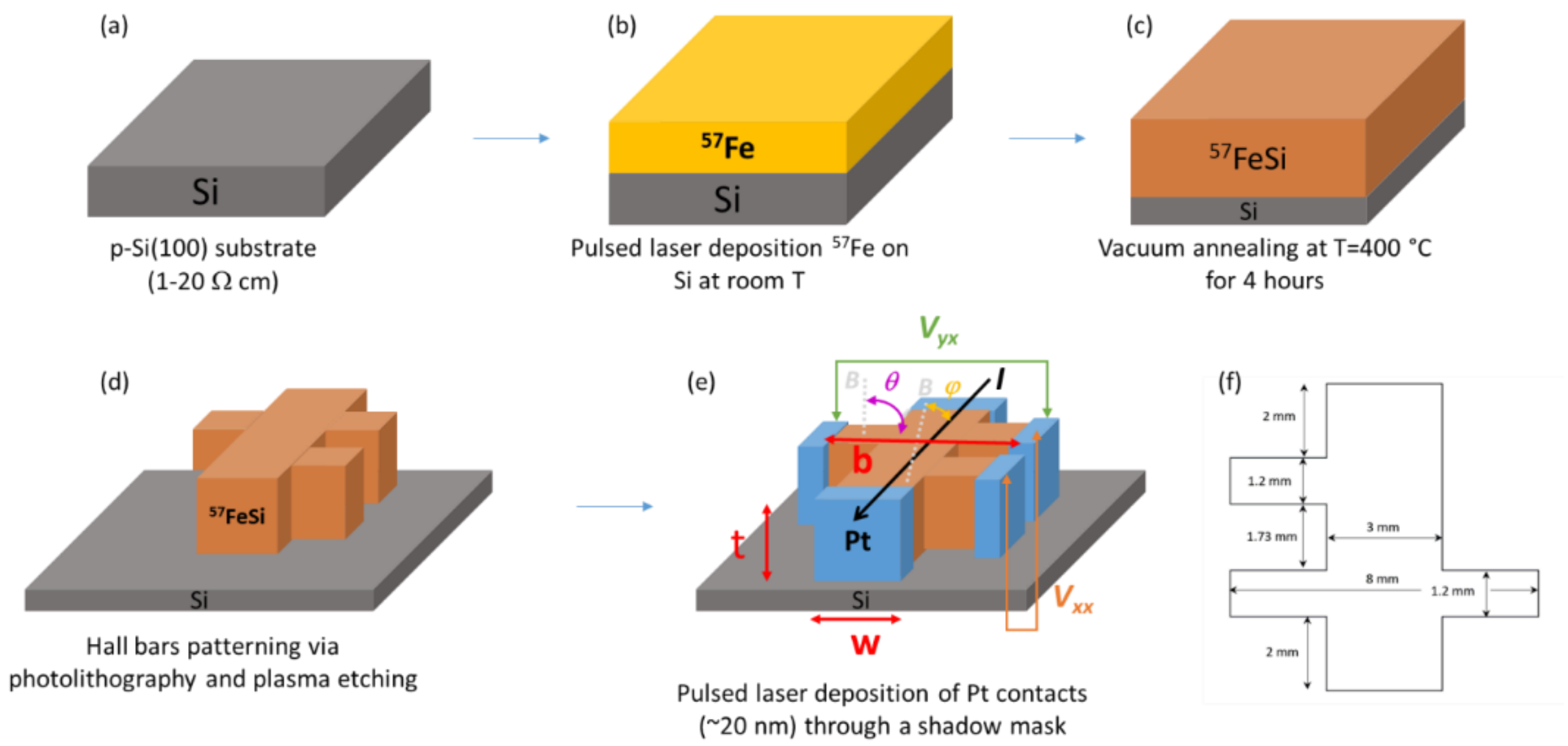

**Figure S1.** (a-c) Fabrication process of a FeSi thin film via PLD of Fe followed by solid-phase reaction carried out through vacuum annealing. (d-f) Hall bar patterning of a FeSi thin film with PLD of Pt contacts. Wide side plates (1.2 mm) in the pattern (f) are designed to switch between the pairs of current and potential probes in van der Pauw commuting scheme.

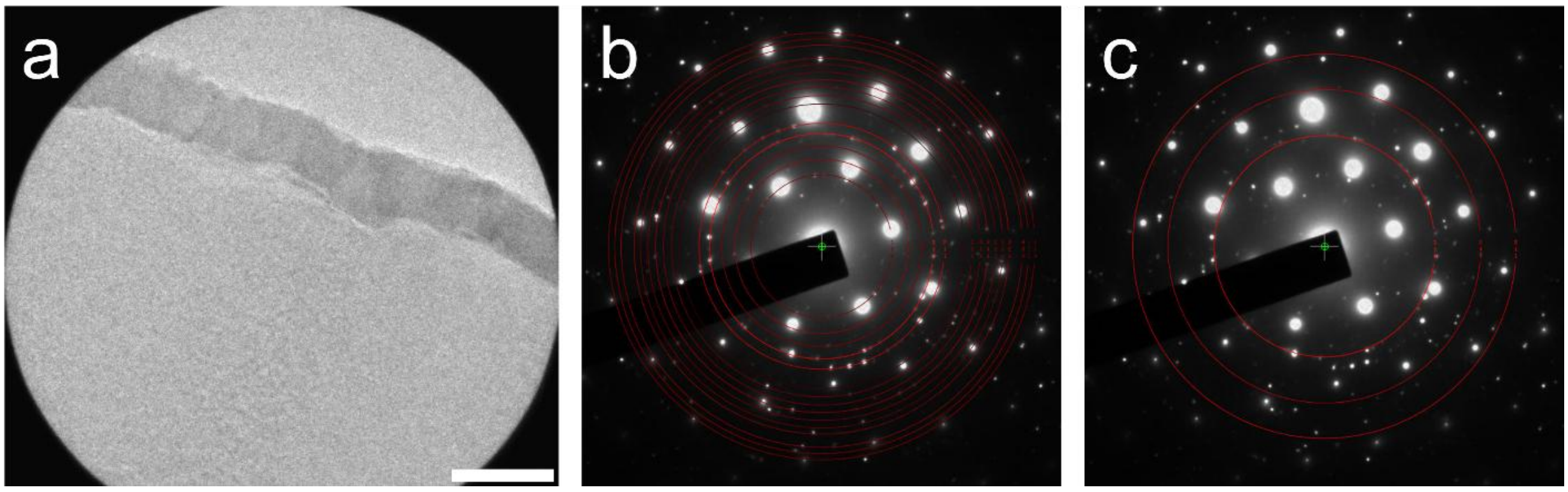

**Figure S2.** Selected-area electron diffraction analysis (SAED). (a) TEM image of a $^{57}$FeSi thin film on Si substrate, used for SAED. Scale bar is 100 nm. (b) SAED pattern of the area shown in (a) with diffraction circles, corresponding to FeSi B20 crystal lattice, superimposed. (c) SAED pattern of the area shown in (a) with diffraction circles, corresponding to α-Fe crystal lattice, superimposed.

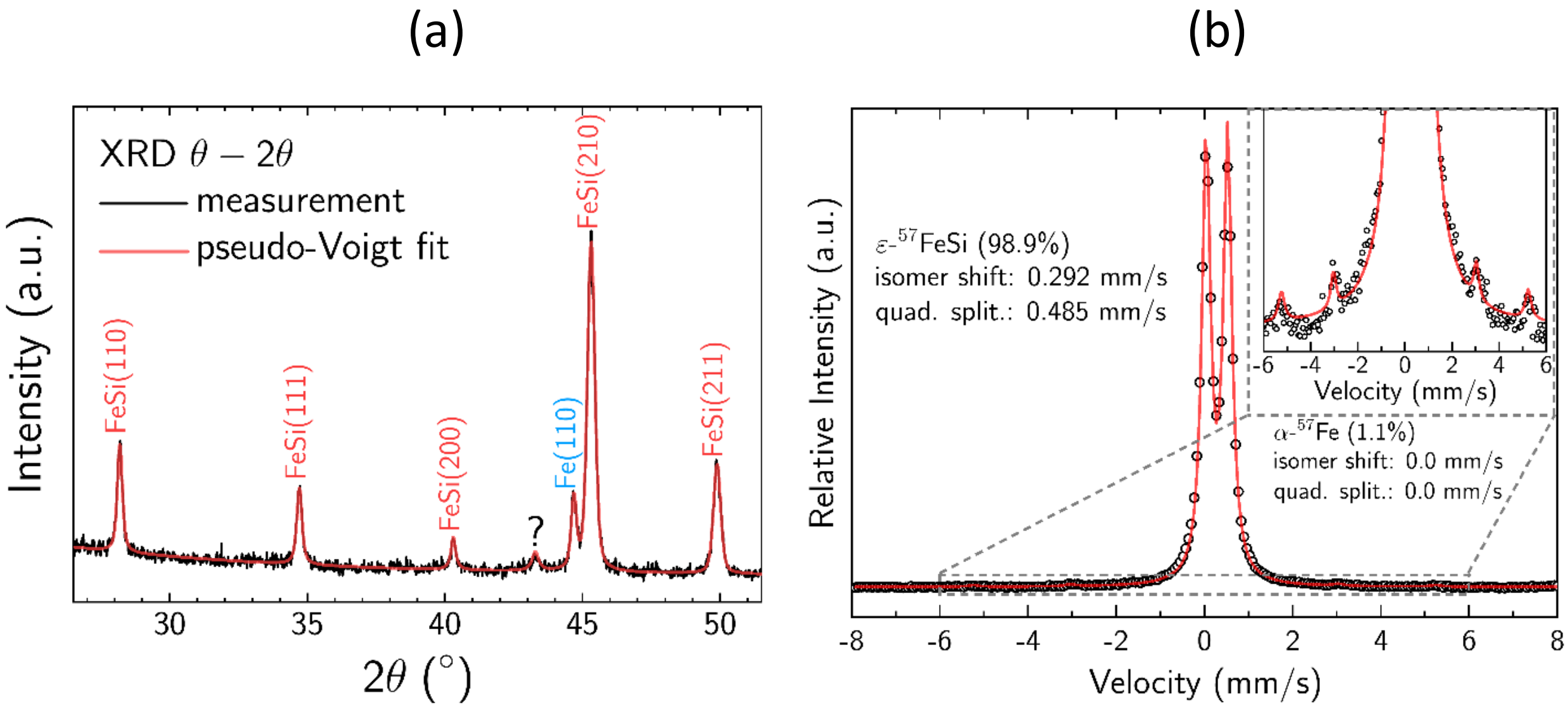

**Figure S3.** (a) X-ray diffraction data and (b) CEMS spectra of $^{57}$FeSi thin films. Tiny satellite peaks in (b) correspond to a residual 1.1 % α-Fe fraction.

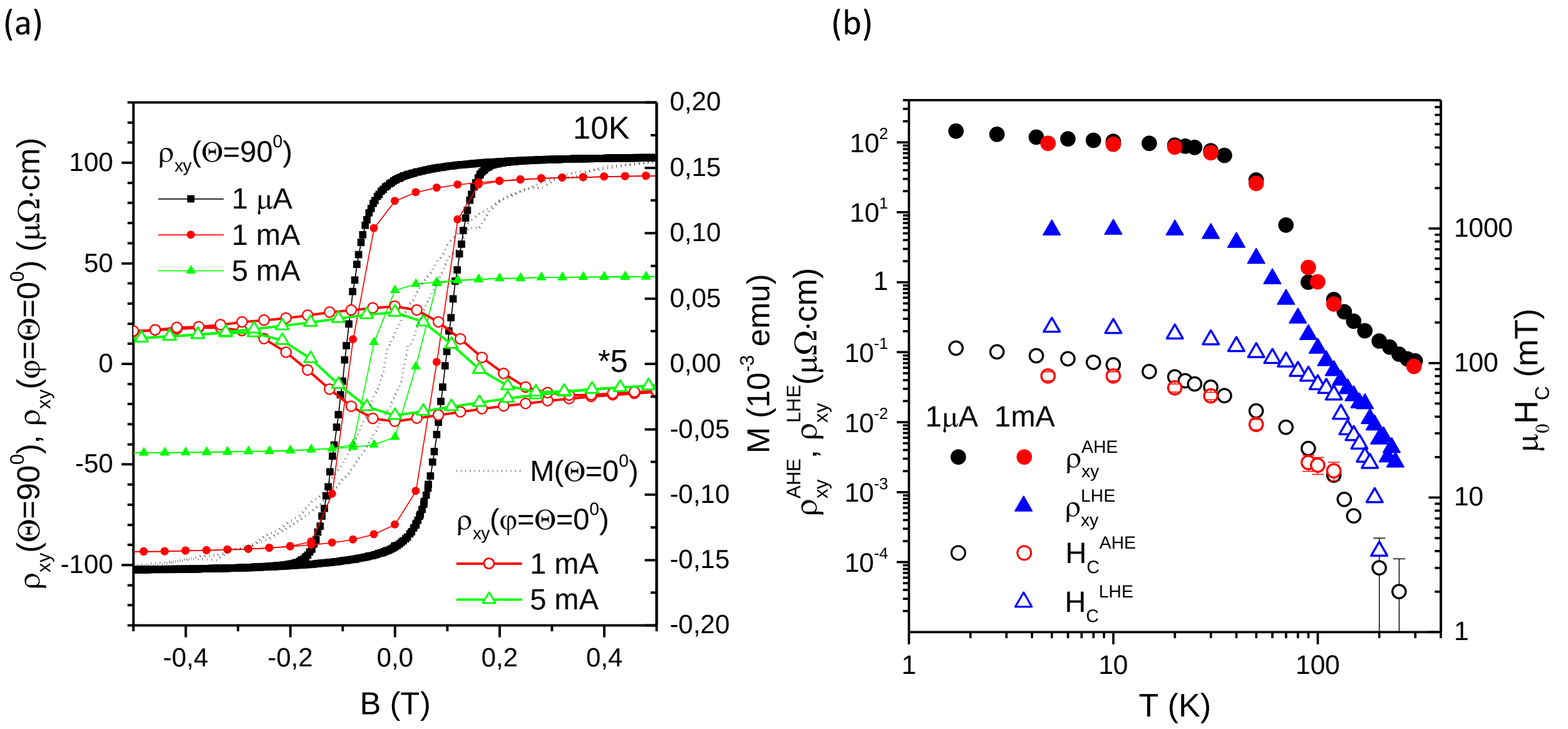

**Figure S4.** (a) Low-field hysteresis loops in transverse ($\Theta=90^0$) and longitudinal ($\varphi=\Theta=0^0$) Hall resistivity $\rho_{xy}$ (scaled by a factor of 5) as measured for 65 nm ε-FeSi thin film at 10K. Black dotted line presents the magnetization M(B) of the FeSi film in magnetic field lying in the plane ($\varphi=\Theta=0^0$).

(b) The amplitudes of anomalous Hall resistivity measured for both transverse ($\rho_{xy}^{AHE}$, $\Theta=90^0$) and longitudinal ($\rho_{xy}^{LHE}$, $\varphi=\Theta=0^0$) directions of magnetic field (closed symbols, left y axis) compared to the corresponding magnetic coercive forces $\mu_0 H_C$ (open symbols, right y axis). The excitation currents applied are shown in the legends.

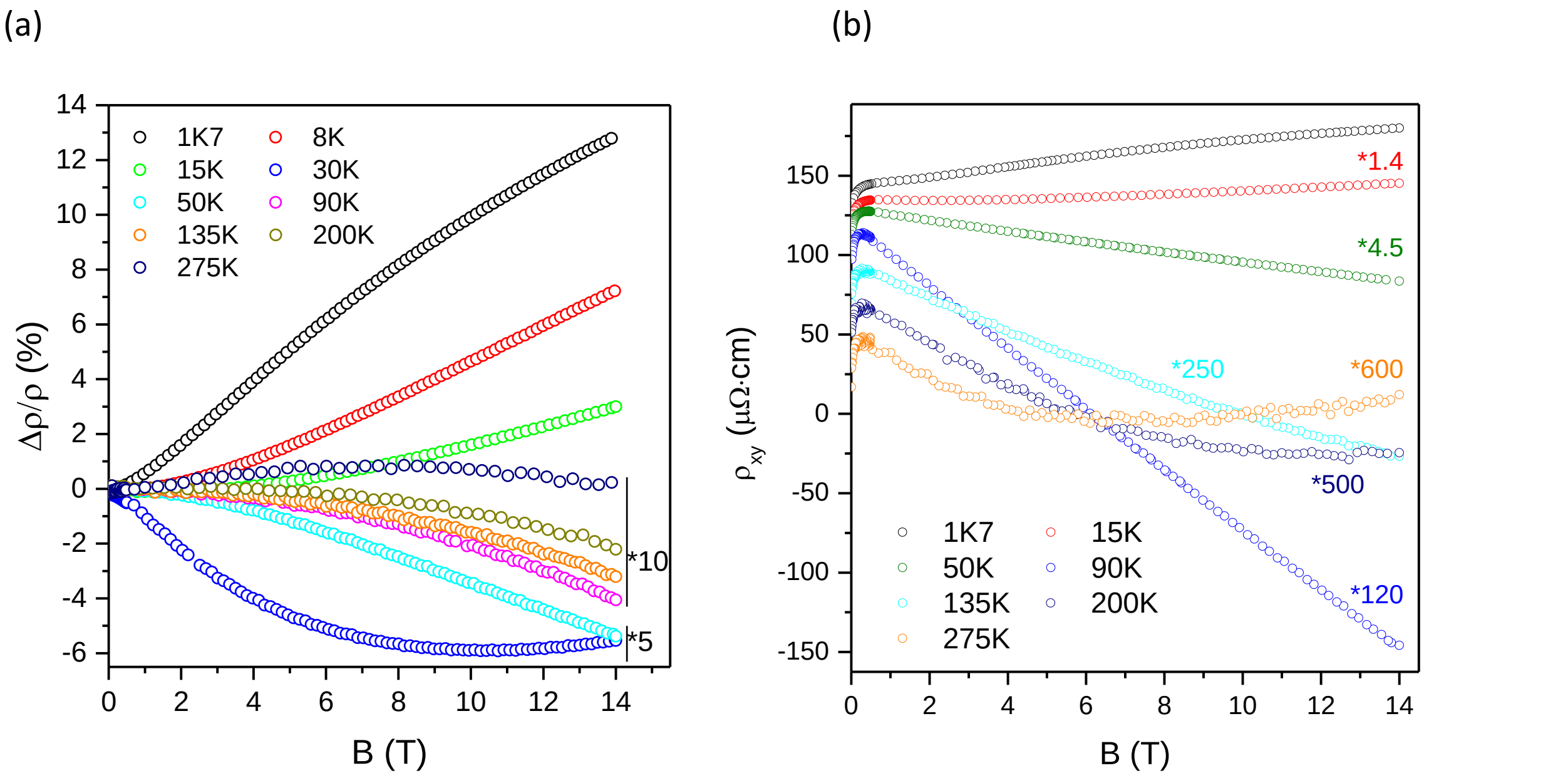

**Figure S5.** Field dependences of (c) measured for 65 nm ε-$^{57}$FeSi thin film at different temperatures between 1.7 K and 300 K. Numbers shown in the plot are scaling factors applied to compare the effects with different amplitudes.